# Estimating the Diffuseness for the Non-Relaxor Type Ferroelectric to Paraelectric Phase Transition in $BaTiO_3$

Prithwiraj Ganguly[a], Prashant Joshi[a], Maneesha Puthiyoth[a], Dilip Sasmal[a], Somaditya Sen[a*]
[a]Department of Physics, Indian Institute of Technology Indore, 453552, India
*Corresponding author; sens@iiti.ac.in

**Abstract**
The normal ferroelectric to paraelectric phase transition in model ferroelectric materials such as $BaTiO_3$ is typically characterized by a sharp, well-defined dielectric constant peak at a specific transition temperature. However, under certain modifications of the parent material, this transition can become diffuse over a broad range of temperatures. This has garnered significant research attention over the past few decades, primarily because of its intriguing and not yet fully understood physical properties. The parameters developed to measure the diffuseness are also ambiguous. In this work, an investigation has been conducted to understand the transition dynamics of the non-relaxor ferroelectric systems in the temperature interval over which the diffuse phase transition occurs. This is achieved by modelling the dielectric response phenomenologically, using a distribution of local transition temperatures. Moreover, a temperature-dependent differential analysis is introduced, enabling the clear demarcation of distinct dielectric regimes and providing enhanced insight into the evolution of the phase transition. It is strongly recommended that the degree of diffuseness 'γ' parameter is skeptical in many senses. Hence, following the establishment, a simple yet effective new measure of diffuseness is being proposed, offering a more accurate and physically meaningful estimation of the diffuse phase transition.

**Introduction**
Phase transitions in a material occur when the thermodynamic free energy of a system exhibits a non-analytic behaviour as a function of an order parameter, for some choice of thermodynamic variables (like temperature, pressure) [1]. The Ferro-Electric (FE) to Para-Electric (PE) phase transition is a classic example of such a transition, characterized by a change in the polarization state of the material. The presence of domains in the FE phase causes the existence of a net spontaneous polarization ($P_s$) in the system even after switching off the external field completely. With an increase in temperature, the $P_s$ vanishes in the PE phase. Barium titanate ($BaTiO_3$ or BTO), with a simple perovskite structure, is a well-known example of FE material. The distorted $TiO_6$ octahedra give rise to a net resultant dipole moment in the non-cubic phases of BTO [2]. However, with increase in temperature, it undergoes a succession of structural phase transitions, in the sequence, Rhombohedral (R3m) (T<183K)→ Orthorhombic (Amm2) (183K<T<278K)→ Tetragonal (P4mm) (278K<T<393K)→ Cubic (Pm-3m) (T>393K) phases [2,3]. Among them, the FE to PE transition is accompanied by the polar tetragonal to highly symmetric cubic structural phase transformation. Here, at elevated temperatures, the thermal randomization disrupts the long-range cooperative alignment of the dipoles, and hence the $P_s$.
In the pristine polycrystalline BTO, although there are different domains, they are of the same dielectric nature. Thus, here the Normal FE to PE Phase Transition (NPT) is characterized by a sharp, well-defined dielectric constant ($\varepsilon'_r$) peak at the transition temperature ($T_c$), with minimal frequency dependence, and it follows the Curie-Weiss law (CW law) immediately after the transition:

$$\frac{1}{\varepsilon'_r(T)} = \frac{(T-T_0)}{C} \tag{1}$$

where 'C' and '$T_0$' are the Curie constant and Curie temperature, respectively. Note that, as the transition is of first order, $T_0<T_c$ [4].

Unlike the pristine one, with chemical modifications in BTO, the system becomes compositionally disordered with the presence of a randomized distribution of various cations among the equivalent lattice sites and in different amounts. These give rise to polar regions of different dielectric nature with different $T_c$ s. Thus, in contrast to NPT, the phase transition gets extended over a wide temperature interval (Curie range) around a maximum value of $\varepsilon'_r$ ($\varepsilon'_m$) near $T_m$ [5,6]. This is known as Diffuse Phase Transition (DPT). Unlike normal FEs, here the $P_s$ is non-zero even for $T>T_m$, and gradually decreases with temperature, extending well above it [7,8]. In certain cases, the temperature $T_m$ shifts prominently towards higher values with an increase in frequency, showing a pronounced dispersive feature before transition. These materials are classified as Relaxor FEs (RFE) [6,8,9]. This can be explained by describing the formation of Polar Nano Regions (PNRs) dispersed in a disordered matrix [7,10]. Many theoretical models have been developed so far to explain the origin of this kind of behaviour [7,11,12]. However, in some cases, though the dielectric peak is diffused and shows frequency dispersion, the $T_m$ does not get shifted with a change in frequency [13]. Materials showing such an incomplete DPT nature, commonly referred to as FE-DPTs [14]. Many such materials have been discovered so far, through targeted chemical modification of the prominent FEs lattices [15–18]. The difference in the frequency response in RFEs and FE-DPTs causes the presence of PNRs to be sceptical, in the latter. However, this could be explained qualitatively in terms of different PNR sizes and their relaxation time. The PNR sizes of the RFEs are small, resulting in a relaxation frequency that falls in the accessible measurement frequency range [13]. However, the FE-DPT is less compositionally disordered, and the size of the newly grown polar regions is so huge that they take a longer time to relax, making the relaxation frequencies too small to interfere with the accessible measurement frequency range [14]. It is observed that the FE-DPT samples have the scale of disorder (for low doping) in the crossover from pure FE to the relaxor nature. In that phase, it still shows proper domain formation [19]. Despite extensive investigations, a clear understanding of the microscopic mechanisms responsible for this behaviour remains elusive to date. The width of the temperature range where the transitions are taking place is a very effective parameter for DPT ferroelectrics. On distinguishing that, it can be effectively considered as the measure of diffuseness. Here, in this work, such an approach has been taken, considering the pros and cons of all the pre-existing models. The investigation focuses on three representative BTO-based FE-DPT systems: Ni-doped BTO [BNTO] and solid solutions of BTO with $CaMnO_3$ [BTO-CMO] and $LaFeO_3$ [BTO-LFO], along with the pristine one. The frequency-dependent plots show no frequency-dependent shift of the $T_m$ value, proving their FE-DPT nature [20,21]. Hence, the samples do not satisfy the Vogel-Fulcher law, which is the concrete, observable evidence of relaxor behavior and the dynamic glass-like freezing [22]. This motivates us to look for new physical insights for these special kinds of FEs. The compositions [BNTO, BTO-CMO, and BTO-LFO] provide a wider platform to understand the versatility of the nature of transition. It is essential to begin by analysing the existing models and identifying their limitations to formulate a more accurate and comprehensive version.

**Problems with the DPT analysis: Different existing models and their drawbacks**

One of the well-known ways to understand the DPT nature is to plot the inverse of $\varepsilon'_r$ with temperature and investigate the deviation from the CW law above $T_m$ [23]. For the pure BTO, the linearity extends up to $T_c = T_m$ [Supplemental Material: Figure 2(a)]. Whereas the DPT systems show a significant deviation from the linear nature over a wide temperature range. The temperature after which the $1/\varepsilon'_r$ plot is linear is described in literature as the Deviation temperature (referred hereafter as $T_{dev}$) [23]. For RFEs, $T_{dev}$ lies at a very high temperature from $T_m$, and so the linearity. Before that, it follows a square temperature dependence instead. Smolensky, for the first time, proposed a semi-quantitative treatment based on a distribution of uncorrelated local polar regions with different $T_c$ s. Considering this distribution to be Gaussian [6]:

$$f(T_c) = \frac{1}{\sqrt{2\pi\delta^2}} \exp\left(-\frac{(T_c - <T_c>)^2}{2\delta^2}\right) \quad (2)$$

where 'δ' and $<T_c>$ represent the standard deviation and the mean value of the distribution.
From Debye's theory, the real ($\varepsilon'_r$) and the imaginary ($\varepsilon''_r$) components of the dielectric constant ($\varepsilon_r$) can be expressed as $\varepsilon_r = \varepsilon'_r + i\varepsilon''_r$ [6], with

$$\varepsilon'_r = \varepsilon_\infty + \frac{8\pi n\mu^2 V_0}{7k_B T(1+\omega^2\tau^2)}$$

$$\varepsilon''_r = \varepsilon_\infty + \frac{8\pi n\mu^2 V_0 \omega\tau}{7k_B T(1+\omega^2\tau^2)} \quad (3)$$

The static dielectric constant is approximated here following the Langevin model for the polar systems with non-interactive dipoles with certain correction factors.
An approximate quadratic relation between $\varepsilon'_r$ and temperature T can be readily obtained by incorporating the contribution from the distribution in equation 3, in terms of a power series expansion as [6]:

$$\frac{1}{\varepsilon'_r(T)} = \frac{1}{\varepsilon'_m}\left(1 + \frac{(T - <T_c>)^2}{2\delta^2}\right) \quad (4)$$

Furthermore, the standard deviation 'δ' is considered to be the measure of diffuseness. It can be obtained by fitting the $\varepsilon'_r$ vs T graph with the above equation for $T>T_m$. In general, $<T_c>$ is considered to be equal to $T_m$ for low-frequency measurement [6], which is not a necessary condition in general. Though this model works quite well for complete RFEs, the disparity between this model and the experimental data is quite high for FE-DPTs. In many cases, the above distribution provides a different value of maxima of $\varepsilon'_r$ and at a different temperature than $T_m$ from the fitting [13]. More importantly, it should be noted that, within this consideration, $\varepsilon_m'$ itself becomes temperature dependent—an aspect that, to the best of our knowledge, has not been reported elsewhere. Therefore, a new empirical power relation was proposed by Uchino and Nomura by replacing the power of (T - $<T_c>$) with a parameter called the degree of diffuseness 'γ' [24]. The modified empirical relation between $\varepsilon'_r$ and T becomes,

$$\frac{1}{\varepsilon'_r(T)} = \frac{1}{\varepsilon'_m}\left(1 + \frac{(T - <T_c>)^\gamma}{2\delta^2}\right) \quad (5)$$

It is to be noted that the diffusivity (δ) and the degree of diffuseness (γ) for a DPT are two different parameters representing two fundamentally different ideas. For $T>T_m$, γ has a value of 1 for normal ferroelectrics and 2 for complete DPTs i.e., RFEs. For FE-DPTs, $1 < \gamma < 2$. So, 'γ' represents how much the sample is deviating from the ordinary CW law and becoming a classical relaxor. Note that, for many systems, the value of 'γ' is obtained from the slope of the linearly fitted curve of $\ln(1/\varepsilon'_r - 1/\varepsilon'_m)$ vs $\ln(T-T_m)$ is found to be different in different temperature regions above $T_m$ [25,26], [Supplemental Material: Figure 1]. Another deviation is obtaining $\gamma>2$ values for certain ceramic systems [27] [28]. Moreover, from the Smolensky model, one can say that 'δ' should have the dimension of temperature. However, in Uchino's relation, such can't be said for $1 \le \gamma < 2$. To solve that, another model has been proposed [5].

One of the important drawbacks of all these models is the importance of $T_m$ and the fitting of the data for $T>T_m$, whereas the $T<T_m$ regime gets completely ignored. One needs to investigate the transition dynamics in the temperature range near the maxima to account for the DPT. Lobo et al. [29] has suggested a relation over a temperature range by taking the convolution of $\varepsilon'_r$ with the distribution of $T_c$ s. Nevertheless, this model has also not been found to be suitable for the entire range of temperatures for BTO [29].

On a separate trial, some analysis has been put forward to estimate the diffusion on the basis of recognizing the temperature range where the diffusion process is taking place. Following that, $\Delta T = T_{dev} - T_m$ is often considered to be a useful measure of diffuseness. As this method also depends on a linear fitting, depending on the fitting range, the $T_{dev}$ comes out to be different [13]. Likewise, the other mentioned models also don't take care of the diffusion effect below $T_m$, and so the overall microscopic distribution of $T_c$ s.

In 2010, Uchino et.al introduced a reasonable and effective diffuseness parameter, taking into account the temperature range both below and above $T_m$, based on Smolensky's microscopic composition fluctuation model and Devonshire's phenomenological theory [30]. In this model, the diffuseness degree (D) was defined as the range of temperature where $\varepsilon'_r$ (T) changes most rapidly as a result of the disappearance of the polar microscopic regions as temperature increases. Thus:

$$D = T_{\left(\frac{\partial \varepsilon'_r}{\partial T}\right)_{min}} - T_{\left(\frac{\partial \varepsilon'_r}{\partial T}\right)_{max}} \tag{6}$$

'D' was proved to be reasonably effective and has good universality for various ferroelectric materials. However, as only after reaching the $T_{dev}$ the samples start to follow the C-W law, therefore, the complete PE phase is not achieved at the temperature $T_{\left(\frac{\partial \varepsilon'_r}{\partial T}\right)_{min}}$, (which is at a lower temperature than $T_{dev}$), making it physically insignificant [Supplemental Material: Figure 3]. Therefore, this model also faces discrepancies in extracting the experimental fact of defining the actual diffusion range.

**Proposed phenomenological model: Measure of diffuseness and functionalizing**

The experimental temperature-dependent $1/\varepsilon'_r$ plots reflect that, for pure BTO the FE phase reveals a sharply diverging nature for $T \leq T_c = T_m$, following the Landau theory [1]. However, for FE-DPTs, this sharp monotonous decreasing nature ends at a temperature far below $T_m$, beyond which, though it is decreasing, but with opposite concavity. The temperature at which the concavity is changing in the $1/\varepsilon'_r$ plot appears as a point of inflection (POI) (and so in the $\varepsilon'_r$ vs. T). Here, beyond this temperature, unlike the pure FE phase, the rate of decrease of $1/\varepsilon'_r$ starts increasing by bringing the change in concavity from down to up. This suggests that the purely FE nature is compromised near the POI. Note that the value of temperature is the same as $T_{\left(\frac{\partial \varepsilon'_r}{\partial T}\right)_{max}}$, i.e., the temperature where the rate of change of $\varepsilon'_r$(T) also achieves its maximum [Supplemental Material: Figure 3]. Thus, taking the analogy from Uchino's work, this nature can be explained by the emergence of the distribution of $T_c$ s. As probabilistically the $T_c$ of some modified polar regions may correspond to even $T<T_m$, allowing certain polar regions to get converted into non-polar PE phase below $T_m$, affecting the monotonous decreasing feature observed for pure FE. The existing phenomenological models have also been tried to explain this nature, on the basis of the distribution of $T_c$ s [27] [31]. Therefore, it supports the fact that this temperature is a good estimate of the effective onset of DPT. It is therefore referred to hereafter as Onset temperature, $T_{ON}$.

To justify the modified temperature-dependent behaviour of $1/\varepsilon'_r$ near the maxima and the change of the concavity after $T_{ON}$, the analysis can be started from the temperature-dependent Debye's model. Equation 3 can also be expressed as equation 7, taking into account the CW law after the transition [32]:

$$\frac{1}{\varepsilon'_r(T)-\varepsilon'_\infty} = \frac{7k_B(1+\omega^2\tau^2)(T-T_0)}{8\pi n\mu^2 V_0} \quad (7)$$

To proceed further, it is important here to understand that one should not confuse the diffuse (non-sharp) phase transition feature with the frequency-dispersion behaviour. Physically, these are two different characteristics. In relaxor ferroelectrics, the domains are not static; rather, due to pronounced structural disorder and random fluctuations, the system supports the formation of dynamic PNRs. Importantly, the 'distribution of relaxation times' has been proposed there to explain the dynamics of PNRs and the frequency dispersion [33]. The diffuseness of the dielectric peak in relaxors originates from the gradual freezing of PNR dynamics with decreasing temperature, which is just another feature of the consideration, although relaxor physics is not universal [34]. Note that, for FE-DPT samples due to an intermediate amount of disorder, PNRs don't form; rather, it still supports domains (smaller than the pure sample) [19]. Hence, the diffuseness should arise from a different statistical occurrence. As a matter of fact, here it is considered that the sample has a single relaxation time (τ). In general, this is also temperature-dependent and follows the Arrhenius equation [13]. However, as for the FE-DPT systems, the relaxation frequency is very small with respect to the measurement frequency (1 MHz), the variation of the $(1+\omega^2\tau^2)$ becomes almost temperature independent [13,35]. Instead, we consider that the parameter 'n', i.e., number density of dipoles contributing to the PE phase, need not be a constant in these systems. Doping the parent material breaks the global phase transition [36]. In general, the common feature shared by all ferroelectrics with DPT is that they possess compositional variations, structural inhomogeneities, or phase heterogeneities in the physical scale from the micron or submicron range to the atomic level. In fact, physically, the inhomogeneity within ferroelectrics influences the manner in which the materials exhibit ferroelectricity [37]. Transmission Electron Microscopy (TEM) [38] and Piezo-response Force Microscopy (PFM) [39] have provided direct evidence of nanoscale structural heterogeneities affecting ferroelectric properties. In $BaTiO_3$, TEM revealed mixed ferroelectric tetragonal and non-ferroelectric hexagonal phases at the nanoscale, a clear indication of local heterogeneity that influences the distribution of local phase dynamics. Similarly, PFM studies have revealed nanoscale domain structures and their dynamics by studying polarization reversal [39], [40]. Multi-scale tetragonal domains and dynamic ferroelectric switching have been reported, with nanoscale mapping of switchable polarization [41]. These variations in switching behaviour point directly to a distribution of local coercive fields and, consequently, to locally varying phase transition characteristics. These local heterogeneities modify the global response by introducing ferroelectric domains with distinct dielectric characteristics. Since the critical transition temperature is associated with the dipole correlation strength—which is inherently non-uniform in this case—it is a good assumption to describe the system using a distribution of Tc values, i.e., $f(T_c)$ [42]. Hence, only those domains that get enough energy to surpass the barrier having $T_c \leq T$, become available to the PE phase. However, the polar region with $T_c \geq T$ has a null contribution. As the temperature rises, more domains start converting into the PE phase, and more dipole moments are added to it. The assumption of a distribution of $T_c$ corresponds to a mesoscopic description of the heterogeneous doped systems, which emerges from structural disorder [43] and variations in domain configuration [44]. As the temperature rises, more domains start converting into the PE phase, and more dipole moments are added to it. Hence, 'n' is no longer a constant but is a temperature-dependent parameter (n(T)) in the DPT range.

Assuming the distribution $f(T_c)$ as equ.2, the expression of n(T) can be expressed as,

$$n(T) = \frac{N}{\sqrt{2\pi\delta^2}}\int_{-\infty}^{T} exp\left[-\frac{(T_C-<T_C>)^2}{2\delta^2}\right] dT_c \quad (8)$$

Where 'N' is the total no. of dipole moments. Therefore, n(T) is proportional to the Cumulative Distribution Function (CDF) of a Gaussian distribution. As there is no exact closed form of this integral, this can be numerically fitted to the experimental points. However, to make the physical interpretation more explicit, it is a good approximation to take the CDF of a Gaussian distribution as a logistic function with $\lambda = \delta/(\sqrt{3}/\pi) \sim 0.5513\times\delta$ [45],

$$\frac{1}{\varepsilon_r'(T)-\varepsilon_\infty'} = \frac{7k_B\sqrt{2\pi\delta^2}(1+\omega^2\tau^2)(T-T_0)}{8\pi\mu^2 V_0 N}\left[1+\exp\left(-\frac{(T-<T_C>)}{\lambda}\right)\right] \tag{9}$$

Thus, from equ.9, for $\varepsilon_r' \gg \varepsilon_\infty'$ near the transition, the temperature dependence comes out to be,

$$\frac{1}{\varepsilon_r'(T)} = \frac{(T-T_0)}{C'}\left[1+\exp\left(-\frac{(T-<T_C>)}{\lambda}\right)\right] \tag{10}$$

with, $C' = \frac{8\pi\mu^2 V_0 N}{7k_B\sqrt{2\pi\delta^2}(1+\omega^2\tau^2)}$ , which is the modified Curie constant.

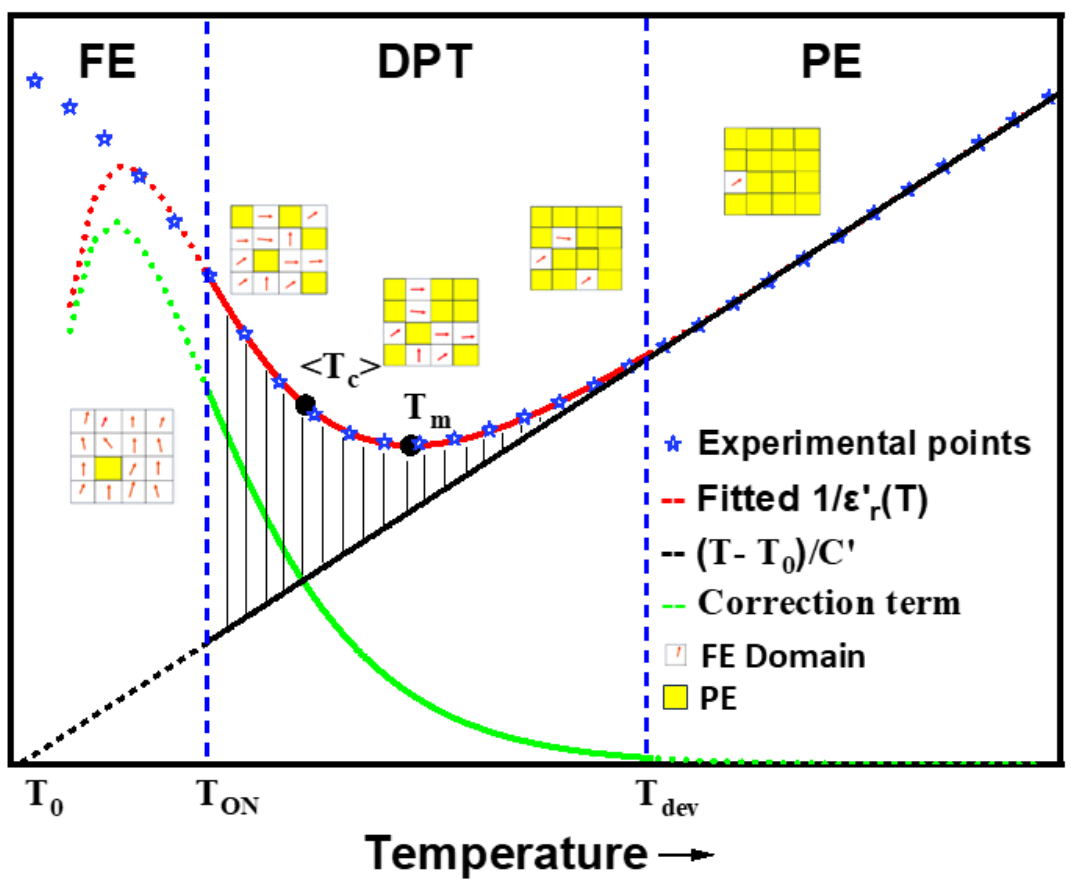

**Figure 1:** Schematic of the proposed model (for BNTO)

The following form reflects the fact that the temperature dependence contains two separate contributions in the $T_{ON}<T<T_{dev}$: firstly, a linear dependence originating from CW law (i.e. pure paraelectric nature) and secondly, an additional correctional contribution from n(T), causing the deviation (i.e. the gradual transition from FE to PE phase) [Figure 1, shown as a shaded region]. The correction term shows maximum contribution at lower temperatures and reveals a nearly decaying nature from temperature, $T_{ON}$ to $T_{dev}$. Near the $T_{dev}$ as all the domains get converted into the PE phase, the 'n(T)' gets saturated, making the correctional contribution just a constant scaling factor. So, after this temperature, the $\varepsilon'_r$ starts to follow ordinary CW law.

The experimental data were fitted with this new expression of $1/\varepsilon'_r(T)$, and the results are absolutely in place [Supplemental Material: Figure 2]. This modified temperature dependence effectively disrupts the diverging nature of the curve by effectively bringing a POI at $T_{ON}$. Note that the fitted $<T_c>$ is not found to be the same as $T_m$ [Supplemental Material: Table 1]. Hence, as claimed earlier, it proves the insignificance of $T_m$ in diffusion measurement as an origin temperature. Furthermore, this modified formulation takes care of both the below and above temperature regions near the maxima. It is essential to emphasize that the proposed model cannot be extrapolated to the pure ferroelectric (FE) region, i.e., below $T_{ON}$, as this corresponds to a distinct phase governed by different thermodynamic principles. Note that the Gaussian distribution of $T_c$ results in the cumulative distribution of dipoles from the FE to PE converted domains. Such a cumulative distribution multiplied by the temperature-dependent term arising from the Curie-Weiss law will have only a single point of inflection. Hence, although a Gaussian distribution may have two points of inflection (POIs) at $<T_c> \pm \delta$ ($\delta$ is the standard deviation), the actual POI doesn't depend on this Gaussian distribution but depends on the functional form of $1/\varepsilon'_r(T)$. Consequently, in the interval $T_{ON}<T<T_{dev}$, the modified temperature dependence of $1/\varepsilon'_r(T)$ depicts an upward concavity nature, in clear contrast to the downward concavity characteristic of the pure FE region. This particular temperature ($T_{ON}$) of the above change is the POI and the initiation point of different thermodynamic processes. These unambiguous generalities underscore that the non-linear dielectric nature in the DPT region ($T_{ON}<T<T_{dev}$) can be well understood within the framework of diffuse transition by taking care of all the discrepancies of the existing models.

**Proposed Differential Plot**

From the following discussion so far, it is understandable that the gradual transformation of the polar FE domain to the non-polar PE phase is prominently active within the two temperature limits $T_{ON}$ and $T_{dev}$. Here, a differential plot of $\partial/\partial T(1/\varepsilon'_r)$ vs. T is being presented to exactly determine and define these two important temperatures. The general feature of the curve is that it starts with a continuous decreasing nature, followed by a continuous increasing range, and at higher temperatures, it gets saturated to a constant value [Figure 2]. The temperature at which it changes its nature from continuous decrement to continuous increment corresponds to the POI in the $\varepsilon'_r(T)$ vs. T curve, i.e., $T_{ON}$. On the other hand, the saturation region corresponds to the linear temperature dependence from $1/\varepsilon'_r(T)$ i.e., the CW law. Thus, the onset of the saturation region represents the temperature $T_{dev}$. Thereby, the entire temperature range can now be divided into three regimes. The $T>T_{dev}$ range is the complete PE region where $\partial/\partial T(1/\varepsilon'_r)$ is a constant, following the linear temperature dependence of $1/\varepsilon'_r$. The second region is the perfectly FE region for $T<T_{ON}$. Keeping correspondence with the pure FE nature of the pristine sample before the transition, in the $T<T_{ON}$ range, the value of $1/\varepsilon'_r$ is reducing monotonously, and the rate of decrease of $1/\varepsilon'_r$ is also decreasing. Lastly, the temperature region $T_{ON}<T<T_{dev}$ is recognized as the DPT region in which the $f(T_c)$, so the n(T) is active. In the DPT region, the $\partial/\partial T(1/\varepsilon'_r)$ plot continuously increases, following the gradual conversion effect of FE to PE phase.

Thus, this temperature range is proposed over here as the new measure of diffuseness ($\Delta$):

$$\Delta = T_{\left[\frac{\partial}{\partial T}\left(\frac{1}{\varepsilon'_r}\right)\right]_{sat}} - T_{\left[\frac{\partial}{\partial T}\left(\frac{1}{\varepsilon'_r}\right)\right]_{min}}$$

$$= T_{dev} - T_{ON} \tag{11}$$

Note that for the pure BTO sample, the DPT region is the lowest, making $T_m$, $T_c$, $T_{dev}$, and $T_{ON}$ nearly degenerated in the limiting case [Supplemental Material: Figure 4] (Though it is not completely degenerated because of the deviation from ideal crystalline nature in the polycrystalline sample). However, the structural changes mediated by doping increase the DPT region significantly for the FE-DPT samples [Supplementary Data: Figure 4]. One should understand that these are not sharp boundaries but a smeared region, which is getting estimated over here as a boundary point to define the region.

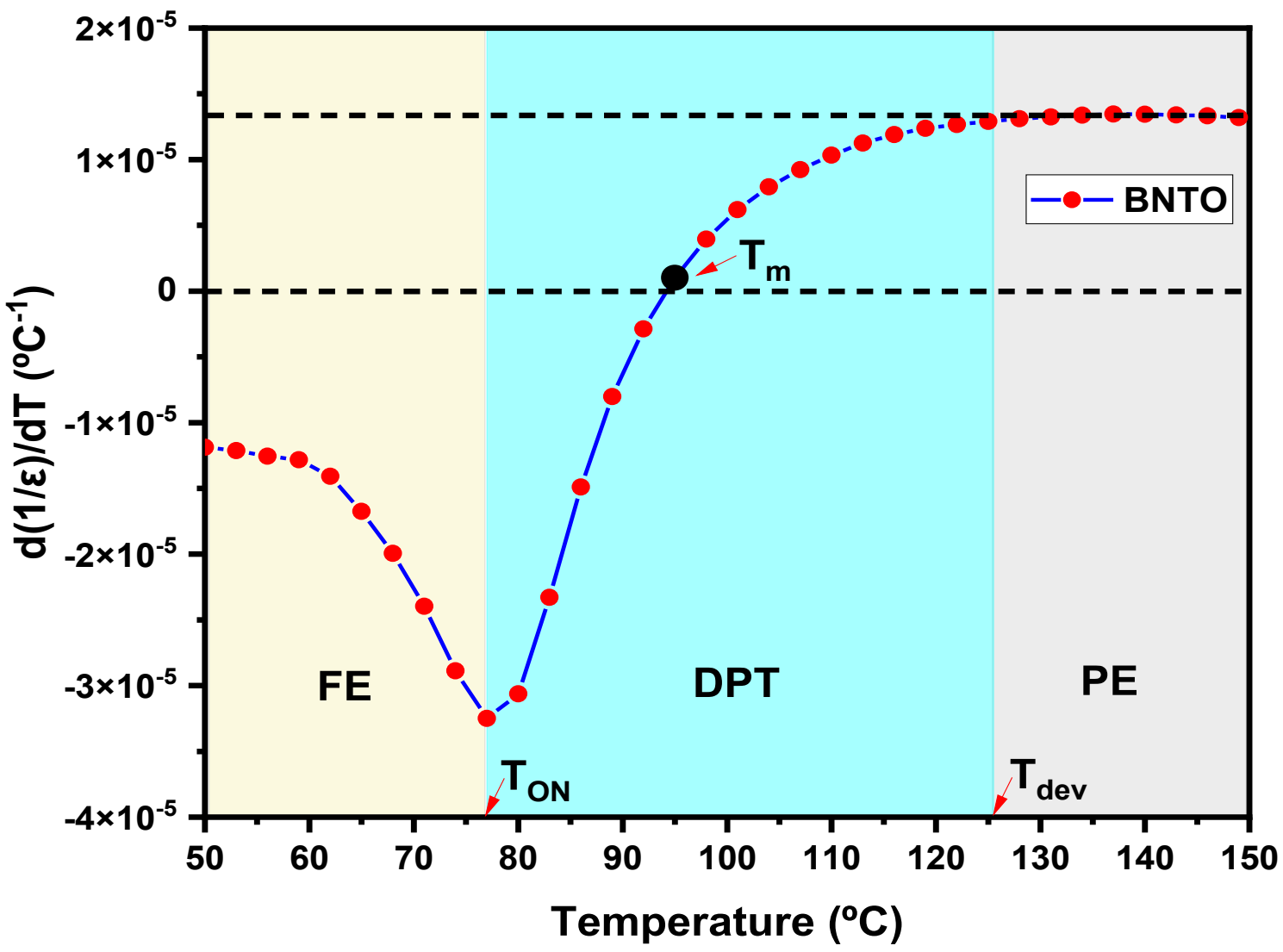

**Figure 2:** Proposed differential plot: $\partial/\partial T(1/\varepsilon_r')$ vs. T

**Conclusion**

A new formalism is presented here to determine the three different temperature regimes in FE-DPT systems by plotting the $\partial/\partial T(1/\varepsilon_r')$ vs T, having a fundamentally different dielectric nature. A theoretical derivation has been provided, following the distribution of transition temperatures, which modifies the temperature dependence of the CW law. It not only explains the functional form of the temperature dependence of $\varepsilon_r'$ near the phase transition, but also explains the discrepancies of the previously proposed models. On demarcating the diffusion region, the temperature range is introduced here as the new measure of diffusion. Nevertheless, the frequency dependence of the diffusion parameter is yet to be studied for both FE-DPT and RFE samples, considering this model. A comparison with the related cubic perovskites indicates that the phase transformation behaviour in BTO is, though not universal, but has many similarities to other FE materials. Thus, in future work, it may be generalized for other diffuse ferroelectric materials too.